\documentclass[aip,apl,reprint,amsmath,amssymb,floatfix]{revtex4-2}
\usepackage{graphicx}
\usepackage[colorlinks, citecolor=blue, linkcolor=blue]{hyperref}
\usepackage{txfonts}
\usepackage{bm}		

\newcommand{\CNN}{Centre de Nanosciences et de Nanotechnologies, CNRS, Universit\'e Paris-Saclay, 91120 Palaiseau, France}
\newcommand{\SPEC}{Service de Physique de l'{\'E}tat Condens{\'e}, CEA, CNRS, Universit\'e Paris-Saclay, 91191 Gif-sur-Yvette, France}
\newcommand{\Muenster}{Institute of Applied Physics, University of Muenster, 48149 Muenster, Germany}

\begin{document}

\title{Stimulated magnon scattering by non-degenerate parametric excitation}

\author{Joo-Von Kim}
\email{joo-von.kim@cnrs.fr}
\affiliation{\CNN}
\author{Hugo Merbouche}
\affiliation{\SPEC}
\affiliation{\Muenster}
\date{16 September 2024}

\begin{abstract}
Parametric spin wave excitation allows studying a variety of nonlinear phenomena, such as magnon scattering. In patterned micro- and nanostructures the magnon spectra is discrete and translational symmetry is broken, which means allowable scattering channels differ from those in continuous films. An example is non-degenerate scattering by which high-power transverse field pumping creates two magnons with distinct frequencies around half the pumping frequency. Through micromagnetics simulations, we show under certain conditions that combining two pumping frequencies generates new magnon modes through a process of stimulated magnon scattering. Such processes are found to depend on the film geometry and sequence of the pumping fields.
\end{abstract}

\maketitle

Nonlinear spin wave dynamics in magnetic thin films and devices has found renewed interest due to potential applications in unconventional computing.~\cite{Nakane:2018jc, Watt:2020di, Papp:2021ga, Korber:2023pr} For neuro-inspired computing, a promising approach involves mapping the reciprocal space ($k$-space) spanned by spin wave eigenmodes to a recurrent neural network, whereby information propagates through the network by nonlinear processes such as magnon scattering. For example, three-magnon scattering in vortex states within micrometer-sized disks has proven effective for pattern recognition tasks, underpinning the nonlinear transformation between input and output magnon spectra in the radio frequency (RF) domain.~\cite{Korber:2023pr}

This result motivates us to enquire whether features of the $k$-space neural network can be harnessed with other systems. Generally, parametric excitation offers a direct means to input information into the spin wave network by targeting specific modes. For example, this can be achieved with parallel pumping,~\cite{Bracher:2011jn, Ulrichs:2011dz, Guo:2014cw, Hwang:2021pe, Heinz:2022pg} although much remains to be explored on how several modes could be populated at once.

\begin{figure}
	\centering\includegraphics[width=8.5cm]{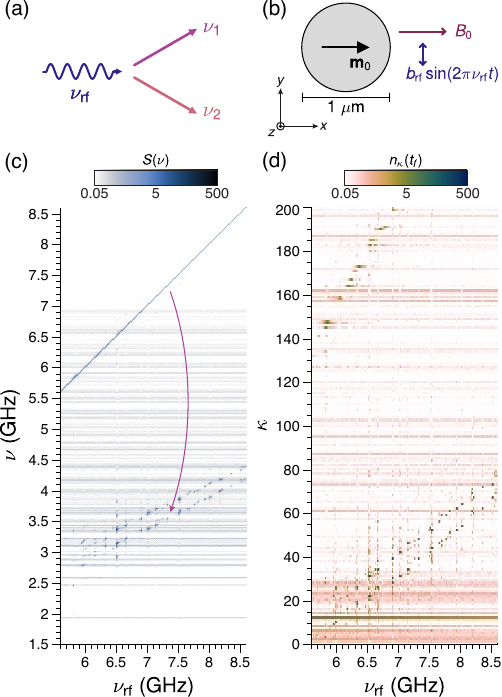}
	\caption{(a) Non-degenerate magnon scattering due to RF field excitation. (b) Disk geometry studied and field orientations considered. (c) Color map of power spectral density of excited spin waves $S(\nu)$ as a function of the frequency of the applied RF field, $\nu_\mathrm{rf}$, for $B_0 = 50$ mT. The arrow indicates the scattering in (a). (d) Population of mode $\kappa$, $n_\kappa$, as a function of  $\nu_\mathrm{rf}$ for the case considered in (c).}
	\label{fig:geometryPSD}
\end{figure}

Inspired by recent experiments on yttrium iron garnet (YIG) disks,~\cite{Merbouche:2024da} we examine using micromagnetics simulations how non-degenerate parametric excitation allows driving several magnon modes simultaneously. Like three-magnon scattering in confined geometries,~\cite{Camley:2014cg, Verba:2021to} transverse pumping with an RF magnetic field at a frequency $\nu_\mathrm{rf}$ can generate pairs of modes $\nu_1$ and $\nu_2$ such that $\nu_\mathrm{rf} = \nu_1 + \nu_2$ with $\nu_1 \neq \nu_2$, as shown in Fig.~\ref{fig:geometryPSD}(a).
Moreover, we find that when two pairs of magnons are pumped to sufficiently high levels by two sinusoidal driving fields with frequencies, $\nu_{\mathrm{rf},A}$ and $\nu_{\mathrm{rf},B}$, stimulated magnon scattering processes can occur. These processes create new scattering channels related to either $\nu_{\mathrm{rf},A}$ or $\nu_{\mathrm{rf},B}$ but are absent when only one of the two are present, suggesting a mechanism to forge new connections between spin wave neurons within the $k$-space paradigm.

Our focus is the system in Fig.~\ref{fig:geometryPSD}(b). It comprises a ferromagnetic thin film disk with a diameter of 1 $\mu$m and a thickness of 50 nm. A static magnetic field, $B_0$, is applied along the $x$ direction and a sinusoidal radio-frequency field, $b_\mathrm{rf}$, along the transverse $y$ direction. $\mathbf{m}_0$ denotes the static magnetization and is mainly aligned along $x$. The time evolution of the magnetization is calculated using the \texttt{MuMax3} code~\cite{Vansteenkiste:2014et}, which uses the finite difference method to solve the Landau-Lifshitz equation with Gilbert damping,
\begin{equation}
	\frac{d \mathbf{m} }{dt} = -|\gamma_0|  \mathbf{m} \times \left( \mathbf{H_{\mathrm{eff}}} + \mathbf{h_{\mathrm{th}}} \right) + \alpha \mathbf{m} \times \frac{d \mathbf{m} }{dt},
	\label{eq:LLG}
\end{equation}
where $\mathbf{m}(\mathbf{r},t)$ is the unit vector representing the magnetization, $\gamma_0 = \mu_0 g \mu_B / \hbar$ is the gyromagnetic constant, and $\alpha$ is the Gilbert damping constant. The effective field, $\mathbf{H}_\mathrm{eff} = -(1/\mu_0M_s)\delta U/\delta \mathbf{m}$, the variational derivative of the total magnetic energy $U$ with respect to the magnetization (with $M_s$ being the saturation magnetization), contains contributions from the Zeeman, nearest-neighbor Heisenberg exchange, and dipole-dipole interactions. Finite temperature is modeled by adding to the effective field a random term $\mathbf{h}_\mathrm{th}$, which has zero mean, $\langle  \mathbf{h}_\mathrm{th} \rangle = 0$, and represents a Gaussian white noise with the spectral properties
\begin{equation}
\langle  h_{\mathrm{th},i}(\mathbf{r},t) h_{\mathrm{th},j}(\mathbf{r}',t') \rangle = \frac{2 \alpha k_B T}{\mu_0 V} \delta_{ij} \delta(\mathbf{r}-\mathbf{r}') \delta(t-t'),
\end{equation}
where $i,j$ represent the different Cartesian components of the field vector, and $V$ is the volume of the unit cell.~\cite{Brown:1963cb} We solve this Langevin problem using an adaptive time-step method.~\cite{Leliaert:2017ci} The disk is discretized using $128 \times 128 \times 1$ finite difference cells. We used an exchange constant of $A = 3.7$ pJ/m, a saturation magnetization of $M_s = 141$ kA/m, and a Gilbert damping constant of $\alpha = 1.5\times 10^{-3}$, which are modeled after recent experiments on YIG thin-film disks.~\cite{Srivastava:2023io}

We study the magnon population dynamics using mode-projection.~\cite{Massouras:2024} This approach involves analyzing the magnetization fluctuations $\delta \mathbf{m}(\mathbf{r}, t)$ about the ground state $\mathbf{m}_0(\mathbf{r})$, $\mathbf{m}(\mathbf{r},t) = \mathbf{m}_0(\mathbf{r}) + \delta \mathbf{m}(\mathbf{r}, t)$, whereby the fluctuations are expressed in terms of normal modes 
\begin{equation}
\delta \mathbf{m}(\mathbf{r}, t) \simeq \sum_{\kappa=1}^{\infty} c_\kappa(t) \bm{\psi}_\kappa(\mathbf{r}) + \mathrm{c.c.},
\label{eq:modefilter}
\end{equation}
where ``c.c.'' means complex conjugate. The vector fields $\bm{\psi}_\kappa(\mathbf{r})$ represent normal mode profiles and $c_\kappa(t)$ the complex mode amplitudes, such that the corresponding mode population is given by $n_\kappa = c^{*}_\kappa c_\kappa$. By projecting out these amplitudes directly in the simulations, we can track the time evolution of individual modes even when the frequency spacing between them is below the resolution allowable by the finite duration of the simulations. With this method, we can unambiguously identify which modes take part in  scattering processes, offering an advantage over frequency-domain techniques since phenomena like frequency pulling and mutual phase locking can complicate spectral analyses.~\cite{Massouras:2024} The normal modes are obtained from diagonalizing a dynamical matrix using the method discussed in Refs.~\onlinecite{dAquino:2009ct, Perna:2022nm, Perna:2022cm} and implemented in the open-source code \texttt{magnum.np}.~\cite{Bruckner:2023mn}

Figure~\ref{fig:geometryPSD}(c) shows a color map of the power spectral density of the spin wave eigenmodes, $S_{\nu}$, as a function of RF field frequency, $\nu_\mathrm{rf}$, for $b_\mathrm{rf} = 5$ mT at $T = 300$ K. For each $\nu_\mathrm{rf}$, we simulate 500 ns during which $c_\kappa(t)$ of the first 200 modes (ordered by their frequencies) are projected out from $\delta \mathbf{m}(\mathbf{r}, t)$, where $\kappa$ is the mode index. The power spectrum is computed for each mode, $S_{\kappa}(\nu) = \int dt \exp{(-i 2 \pi \nu t)} \, c_{\kappa}(t)$, and the sum over these individual spectra, $S_{\kappa}(\nu)$, i.e., $S_{\nu} = \sum_{\kappa=1}^{200}S_{\kappa}(\nu)$ is displayed as a color map in Fig.~\ref{fig:geometryPSD}(c).  Note that only relative changes in $S(\nu)$ are important for subsequent analyses, since their absolute values depend on the normalization constants used and will likely differ to experimentally measured spectra. The horizontal bands between 1.9 and 7.0 GHz represent the thermal spectrum of the 200 modes, while the strong spectral response given by line running between 5.6 and 8.6 GHz represents the direct excitation due to $b_\mathrm{rf}$. The dark bands around several values of $\nu_\mathrm{rf}/2$  correspond to non-degenerate parametric excitation of magnons as schematized in Fig.~\ref{fig:geometryPSD}(a). These only appear within certain frequency intervals because of the discrete nature of eigenmode spectrum.

Figure~\ref{fig:geometryPSD}(d) illustrates the mode-resolved method by presenting the data in Fig.~\ref{fig:geometryPSD}(c) in terms of the mode populations instead. $n_\kappa$ for the first 200 modes are recorded at the end of the simulation, $t=t_f=500$ ns, and plotted using a color map as a function of the input frequency. Like in Fig.~\ref{fig:geometryPSD}(c), we can identify intervals in $\nu_\mathrm{rf}$ in which pairs of modes $(\kappa_1, \kappa_2)$ are excited by the driving field (typically in the range $\kappa < 80$), in addition to modes that directly couple to the RF field in the range $\kappa > 140$. We also observe a dark horizontal line corresponding to $\kappa = 12$, the quasi-uniform mode at this value of $B_0$, which indicates that it is strongly driven non-resonantly by $b_\mathrm{rf}$.

\begin{figure}
	\centering\includegraphics[width=8.5cm]{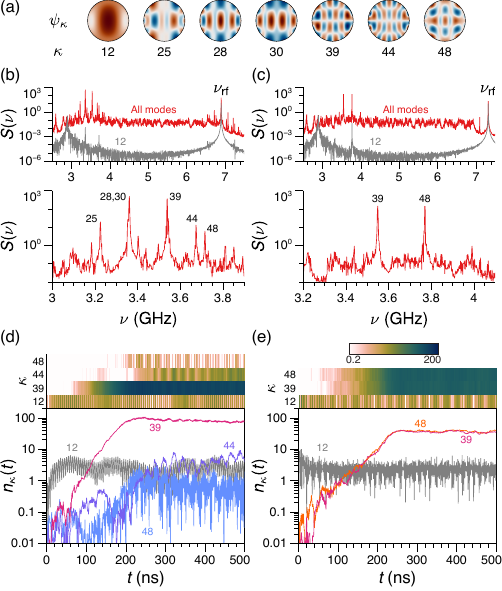}
	\caption{(a) Selected mode profiles, $\psi_\kappa \equiv \bm{\psi}_\kappa \cdot \hat{\mathbf{z}}$. (b) PSD under a driving frequency of $\nu_\mathrm{rf} = 6.90$ GHz. (c) PSD under a driving frequency of $\nu_\mathrm{rf} = 7.32$ GHz. (d) Mode population dynamics corresponding to (b). (e) Mode population dynamics corresponding to (c).}
	\label{fig:nkvst}
\end{figure}

Turning now to the population dynamics of the excited modes, we illustrate the salient features by examining in detail the nonlinear response to $\nu_\mathrm{rf} = 6.90$ GHz, which involves the modes shown in Fig.~\ref{fig:nkvst}(a). 
The power spectral density (PSD) is shown in Fig.~\ref{fig:nkvst}(b) for two frequency ranges of interest. The top panel of Fig.~\ref{fig:nkvst}(b) gives the range encompassing $\nu_\mathrm{rf}$ where peaks of scattered modes can be seen around $\nu_\mathrm{rf}/2$. The gray curve represents the contribution from $\kappa = 12$, which exhibits the strongest response at $\nu_\mathrm{rf}$, confirming non-resonant excitation by the RF field, along with a weaker response around 2.9 GHz, which corresponds to its natural frequency. The bottom panel of Fig.~\ref{fig:nkvst}(b) gives a magnified view of the scattered mode spectra, where peaks are labeled by $\kappa$. $\kappa = 28$ and 30 are excited simultaneously as their natural frequencies are within 1 MHz of each other. The transient dynamics of representative scattered modes, along with $\kappa = 12$, is given in Fig.~\ref{fig:nkvst}(d). $\kappa = 12$ is driven to steady state within tens of nanoseconds, while one of the primary scattered modes, $\kappa = 39$, exhibits an exponential growth over around 200 ns before reaching steady state. Secondary pairs of scattered modes, represented by $\kappa = 44$ and 48, exhibit an incubation time of over 100 ns before growing exponentially and saturating, while remaining orders of magnitude weaker than the primary scattered mode but comparable to $\kappa = 12$. This incubation delay is similar to the behavior observed for parallel pumping in the same geometry.~\cite{Massouras:2024} Above the curves in Fig.~\ref{fig:nkvst}(d), we use a color map to present the same time dependence of $n_\kappa$ on a logarithmic scale. We will rely heavily on this visualization scheme in subsequent figures.

A similar analysis of the scattered mode PSD and population dynamics is presented in Figs.~\ref{fig:nkvst}(c) and \ref{fig:nkvst}(e), respectively, for a driving frequency of $\nu_\mathrm{rf} = 7.32$ GHz. In contrast to the previous case, only one pair of scattered modes is observed, $\kappa = 39$ and 48. As Fig.~\ref{fig:nkvst}(e) shows, their population dynamics and levels mirror one another, as expected for the scattering process in Fig.~\ref{fig:geometryPSD}(a). Note that the populations shown in Figs.~\ref{fig:nkvst}(d) and \ref{fig:nkvst}(e) are orders of magnitude above their thermal levels, which lie in the range of $n = 10^{-3}$ $-$ $10^{-2}$.

\begin{figure}
	\centering\includegraphics[width=6.5cm]{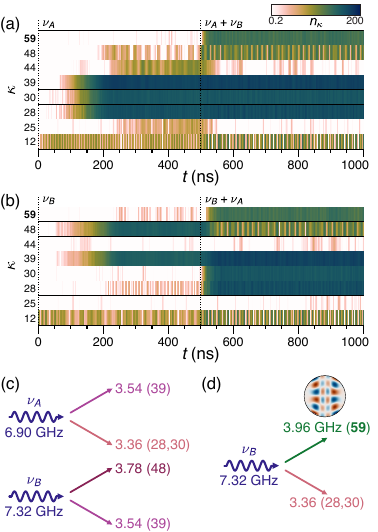}
	\caption{Population dynamics under dual excitation with $\nu_{A} = 6.90$ and $\nu_{B} = 7.32$ GHz. (a)  $\nu_{A} \rightarrow \nu_{A}+\nu_{B}$. (b) $\nu_{B} \rightarrow \nu_{B}+\nu_{A}$. (c) Non-degenerate scattering resulting from $\nu_{A}$ and $\nu_{B}$ separately. (d) Stimulated magnon scattering resulting in $\kappa = 59$, with its profile shown in the inset. Numbers in parenthesis in (c) and (d) refer to the mode index $\kappa$.}
	\label{fig:toggle1}
\end{figure}

Consider now the response to two RF fields with frequencies $\nu_{A}$ and $\nu_{B}$, and the same amplitude of $b_\mathrm{rf} = 5$ mT. We begin with the combination shown in Fig.~\ref{fig:nkvst}, a sequence whereby $\nu_{A} = 6.90$ GHz is applied over 500 ns, followed by $\nu_{A}$ and $\nu_{B} = 7.32$ GHz  together for another 500 ns. We denote this sequence as $\nu_A \rightarrow \nu_A + \nu_B$. The color map of the corresponding population dynamics is given in Fig.~\ref{fig:toggle1}(a).
The expanded view shows the relevant modes involved, but note that a subset of these dynamics was already given in Fig.~\ref{fig:nkvst}(d) for the first 500 ns. When the second pumping field $\nu_B$ is switched on at $t = 500$ ns, a much faster growth toward saturation is seen for $\kappa = 48$, albeit to a lower steady state level compared to that induced by $\nu_B$ alone [Fig.~\ref{fig:nkvst}(e)]. This faster growth is attributed to the other half of the pair, $\kappa= 39$, being already driven to large steady-state levels by $\nu_A$. The presence of both $\nu_{A}$ and $\nu_{B}$ inhibits the secondary pair (25,44), but most interestingly, a new mode is also generated, $\kappa = 59$, which is not associated with the scattering initiated by either $\nu_{A}$ or $\nu_{B}$ individually. Dynamics related to the inverted RF field sequence, $\nu_B \rightarrow \nu_B + \nu_A$, is presented in Fig.~\ref{fig:toggle1}(b). Similarly, the primary scattered modes related to $\nu_{A}$ and $\nu_{B}$ are present, along with the generated mode $\kappa = 59$ when $\nu_{A}$ and $\nu_{B}$ are combined.

To understand the origin of this generated mode, we examine the frequencies of the primary scattered modes resulting from $\nu_{A}$ and $\nu_{B}$ individually, as shown in Fig.~\ref{fig:toggle1}(c), where the figures in parentheses correspond to the mode indices. Because a scattering channel at 3.36 GHz ($\kappa = 28$ and 30) has been opened by $\nu_A$,  parametric excitation with $\nu_{B}$ now results in a new stimulated mode $\kappa = 59$ at a frequency of 3.96 GHz, as shown in Fig.~\ref{fig:toggle1}(d). This is reminiscent of the cross-stimulation process evoked for three-magnon scattering in magnetic vortex states,~\cite{Korber:2023pr, Korber:2020ns} where the presence of a second excitation activates otherwise silent scattering channels.

\begin{figure}
	\centering\includegraphics[width=6.5cm]{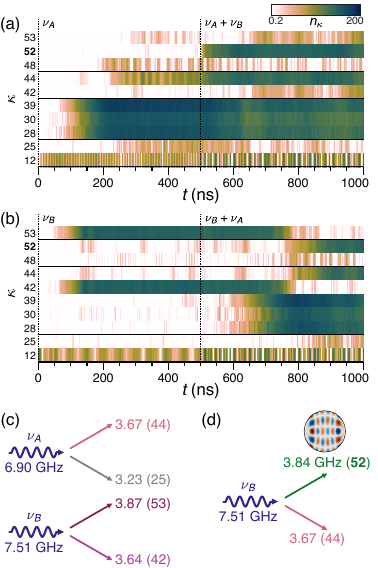}
	\caption{Population dynamics under dual excitation with $\nu_{A} = 6.90$ and $\nu_{B} = 7.51$ GHz. (a)  $\nu_{A} \rightarrow \nu_{A}+\nu_{B}$. (b) $\nu_{B} \rightarrow \nu_{B}+\nu_{A}$. (c) Non-degenerate scattering resulting from $\nu_{A}$ and $\nu_{B}$ separately. (d) Stimulated magnon scattering resulting in $\kappa = 59$, with its profile shown in the inset. Numbers in parenthesis in (c) and (d) refer to the mode index $\kappa$.}
	\label{fig:toggle2}
\end{figure}

Similar stimulated processes also occur for dual excitations that do not share any common scattering channels. An example is given in Fig.~\ref{fig:toggle2} for $\nu_{A} = 6.90$ and $\nu_{B} = 7.51$ GHz. 
While the sequence involves the same $\nu_{A}$ considered previously, the addition of $\nu_{B}$ here results in the amplification of the secondary mode $\kappa = 44$, resulting in the stimulated mode $\kappa = 52$ [Fig.~\ref{fig:toggle2}(a)]. We note that the primary scattered pair corresponding to $\nu_{B}$, (42, 53), is not driven to levels observed when $\nu_{B}$ is applied alone, as seen in Fig.~\ref{fig:toggle2}(b) for the first 500 ns. The importance of the secondary mode $\kappa = 44$ is again apparent in the stimulated generation of $\kappa = 52$ for the $\nu_{B} \rightarrow \nu_{B}+\nu_{A}$ sequence in Fig.~\ref{fig:toggle2}(b), where the population dynamics of $\kappa = 44$ and 52 mirror each other, appearing in unison around 750 ns with a concomitant suppression in the scattered pair (42,53). Analysis of the scattered mode frequencies in Fig.~\ref{fig:toggle2}(c) further confirms that the secondary pair of $\nu_A$ underpins the stimulated mode generation, as illustrated in Fig.~\ref{fig:toggle2}(d).

\begin{figure*}
	\centering\includegraphics[width=12cm]{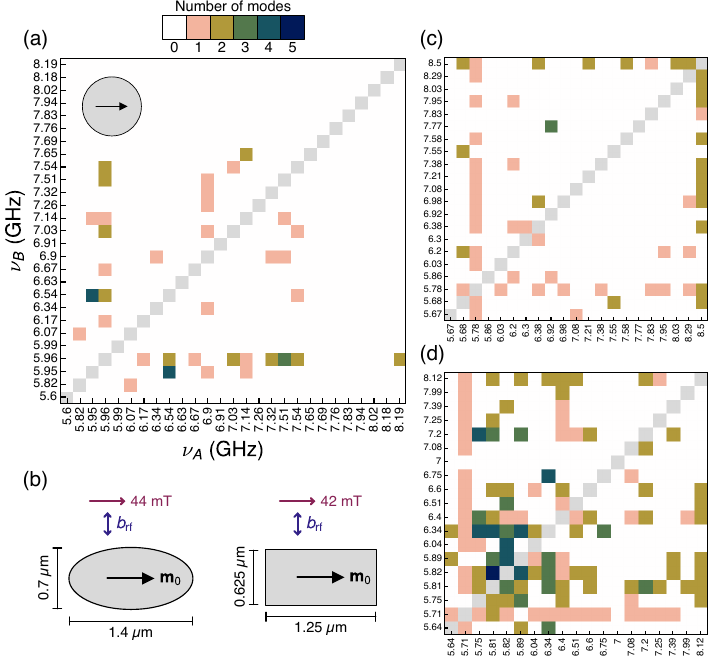}
	\caption{(a) Scattering matrix for dual RF inputs $\nu_A$ and $\nu_B$, with the color code indicating the number of stimulated modes. (b) Other elliptical and rectangular geometries studied. (c) Scattering matrix for elliptical element. (d) Scattering matrix for rectangular element.}
	\label{fig:matrices}
\end{figure*}

Having discussed two stimulated processes in detail, we now comment on their generality in the disk geometry studied. Figure~\ref{fig:matrices}(a) presents a matrix plot in which the number of stimulated modes is shown for the sequence $\nu_{A} \rightarrow \nu_{A}+\nu_{B}$, where the set of $\nu_{A}$ and $\nu_{B}$ are identified from the PSD in Fig.~\ref{fig:geometryPSD}(c).
Some striking features stand out. First, the number of $(\nu_A,\nu_B)$ combinations resulting in stimulated magnon scattering is relatively low, as witnessed by the large swaths of blank entries in the matrix plot. Second, most combinations that do lead to stimulated scattering only produce one generated mode, as shown in Figs.~\ref{fig:toggle1} and \ref{fig:toggle2}. Nevertheless, there are cases in which several modes generated, notably for $(\nu_A,\nu_B) = (5.95,6.54)$ GHz for which four new modes (i.e., two new pairs) are stimulated. Third, the matrix is not symmetric about the diagonal $\nu_A = \nu_B$, indicating that the stimulated generation can be \emph{noncommutative}, i.e., whether $\nu_A$ or $\nu_B$ precedes the dual excitation $\nu_A + \nu_B$ makes a difference to the final outcome.

We now briefly discuss how stimulated magnon scattering is influenced by the film geometry. Similar simulations were performed for an ellipse and a rectangle with a 2:1 aspect ratio, as depicted in Fig.~\ref{fig:matrices}(b). These structures possess the same thickness and similar surface area as the 1-$\mu$m diameter disk studied, while the static field for each shape is chosen such that the corresponding quasi-uniform mode frequency is close to the value of 2.9 GHz found for the disk. Differences in non-degenerate parametric excitation resulting from the same $b_\mathrm{rf} = 5$ mT excitation field are therefore related to the scattering coefficients and frequency spacing between the normal modes.

Figure~\ref{fig:matrices}(c) displays the scattering matrix for the ellipse. Similar features to the disk are seen, such as the variability in number of stimulated modes and the non-commutativity of the excitation pulses, but there are also some striking differences, such as the distribution of $(\nu_A,\nu_B)$ leading to stimulated generation. Specifically, these are concentrated along well-defined rows or columns in the matrix plot, suggesting that only a handful of scattered pairs, i.e., involving 5.78 and 8.50 GHz in particular, are associated with silent channels most amenable to cross-stimulation. Finally, Fig.~\ref{fig:matrices}(d) presents the scattering matrix for the rectangle. A striking difference here is the large number of combinations that result in stimulated generation. These results show that stimulated magnon scattering can appear in a wide variety of geometries, but its prevalence likely depends on details related to the spectrum of the normal modes and coupling efficiency to the driving fields.

Based on these results, we envisage the following avenues for future study. From the theoretical perspective, it would be interesting to explore whether the stimulated processes relate to three-magnon splitting of the off-resonantly driven uniform mode, or to four-magnon processes involving recombination of parametrically-excited modes. This could be quantified by examining coupled equations of motion for the mode amplitudes, which could be constructed from spin wave theory.~\cite{Perna:2022nm, Perna:2022cm}
In addition, evaluation of relevant overlap integrals would also allows us to ascertain the selection rules involved in the scattering process.~\cite{Etesamirad:2023cs} 
 From the experimental perspective, the prediction of qualitatively different scattering matrices for different geometrical shapes could be tested directly using micro-focus Brillouin light scattering, which has already been employed to reveal non-degenerate parametric excitation in thin-film disks.~\cite{Merbouche:2024da} On a broader level, such stimulated processes could also be investigated for nonuniform magnetic ground states such as vortices.

\begin{acknowledgments}
We thank M. Massouras and G. de Loubens for fruitful discussions. This work was supported by the Horizon 2020 Framework Programme of the European Commission under Contract No. 899646 (k-Net).
\end{acknowledgments}

\section*{Author Declarations}
\subsection*{Conflict of Interest}
The authors have no conflicts to disclose.

\subsection*{Author Contributions}
\textbf{Joo-Von Kim:} Conceptualization (equal); Data curation (lead); Formal analysis (lead); Funding acquisition (lead); Investigation (lead); Methodology (lead); Software (lead); Validation (equal); Visualization (lead); Writing – original draft (lead); Writing – review \& editing (equal). \textbf{Hugo Merbouche:} Conceptualization (equal); Formal analysis (supporting); Investigation (supporting); Methodology (supporting); Software (supporting); Validation (equal); Writing – review \& editing (equal).

\section*{Data Availability}
The data that support the findings of this study are openly available in Zenodo at \url{https://doi.org/10.5281/zenodo.13711262} and at \url{https://doi.org/10.5281/zenodo.8409685}, Ref.~\onlinecite{Crameri:2018} and Ref.~\onlinecite{Crameri:2020tm}.

\section*{References}
\bibliography{articles}

\end{document}